\documentclass[
    ,final            
  ]
  {aipproc}

\layoutstyle{6x9}

\begin{document}

\title{The rotation-magnetic field relation}


\author{Ansgar Reiners}{
  address={Universit\"at G\"ottingen}
}

\author{Alexander Scholz}{
  address={University of St Andrews}
}

\author{Jochen Eisl\"offel}{
  address={Th\"uringer Landessternwarte Tautenburg}
}

\author{Gregg Hallinan}{
  address={National University of Ireland Galway}
}

\author{Edo Berger}{
  address={Harvard-Smithsonian Center for Astrophysics} 
}

\author{Matthew Browning}{
  address={University of California at Berkeley}
  ,altaddress={The University of Chicago} 
}

\author{Jonathan Irwin}{
  address={Harvard-Smithsonian Center for Astrophysics}
}

\author{Manfred K\"uker}{
  address={Astrophysikalisches Institut Potsdam}
}

\author{Sean Matt}{
  address={The University of Virginia; Currently at NASA Ames Research Center}
}

\begin{abstract}
  Today, the generation of magnetic fields in solar-type stars and its
  relation to activity and rotation can coherently be explained,
  although it is certainly not understood in its entirety.  Rotation
  facilitates the generation of magnetic flux that couples to the
  stellar wind, slowing down the star.  There are still many open
  questions, particularly at early phases (young age), and at very low
  mass.  It is vexing that rotational braking becomes inefficient at
  the threshold to fully convective interiors, although no threshold
  in magnetic activity is seen, and the generation of large scale
  magnetic fields is still possible for fully convective stars.  This
  article briefly outlines our current understanding of the
  rotation-magnetic field relation.
\end{abstract}

\maketitle


\section{Introduction}

The rotational evolution of stars is the result of the complex
interaction of several fundamental processes. First, the molecular
cloud contracts conserving initial angular momentum spinning up the
central object. Angular momentum can be stored in a disc, which may
brake the rotation of the central object. After the disc is
dissipated, the star can contract reaching the highest rotation rate
after several ten million years of lifetime. Solar-type stars, i.e.,
stars with convective envelopes, start generating magnetic fields that
couple to the stellar wind. The interaction between charged particles
in the wind and the magnetic field generates a torque braking the
star's rotation. In the case of the Sun, braking has led to a rotation
rate of about 1 revolution every month.

According to the rotation-activity relation, rapidly rotating stars
produce strong magnetic fields generating a strong magnetic torque
that brakes the star. This leads to slower rotation, which in turn
weakens the magnetic field production and braking is weakening, too.
At young ages in open clusters, we observe rapidly rotating, very
active stars, while the (single) field stars generally are slowly
rotating and only weakly active. This means that in principle rotation
and activity can tell about the age of a star
\citep[e.g.,][]{Barnes03, Barnes07}.

The connection between rotation, stellar wind, magnetic fields, and
magnetic activity is reviewed in this splinter session summary. First,
we give an overview about the current picture on rotation in both
clusters and the field, i.e., in young and old stars. Next, we discuss
results from direct and indirect magnetic field measurements and their
connection to stellar wind theory. In the last part, we give a summary
on the theoretical work on magnetic field generation through stellar
dynamos. Low-mass stars and in particular the regime where stars
become completely convective currently present a rather puzzling
picture of the connection between magnetism, activity, and rotation.
Thus, low-mass stars are in the focus of our summary.

\section{Rotation}

\subsection{Young objects}

The net effect of early stellar evolution and disc-coupling is that a
star has approximately constant angular velocity for the first few Myr
while it is still coupled the disc. Then it spins up rapidly once the
disc dissipates, reaching maximum rotational velocity close to when it
arrives on the zero age main sequence, followed by a gradual decay due
to the stellar wind, lasting for the remainder of the main sequence
lifetime.

\begin{figure}
\includegraphics[height=\textwidth,angle=270]{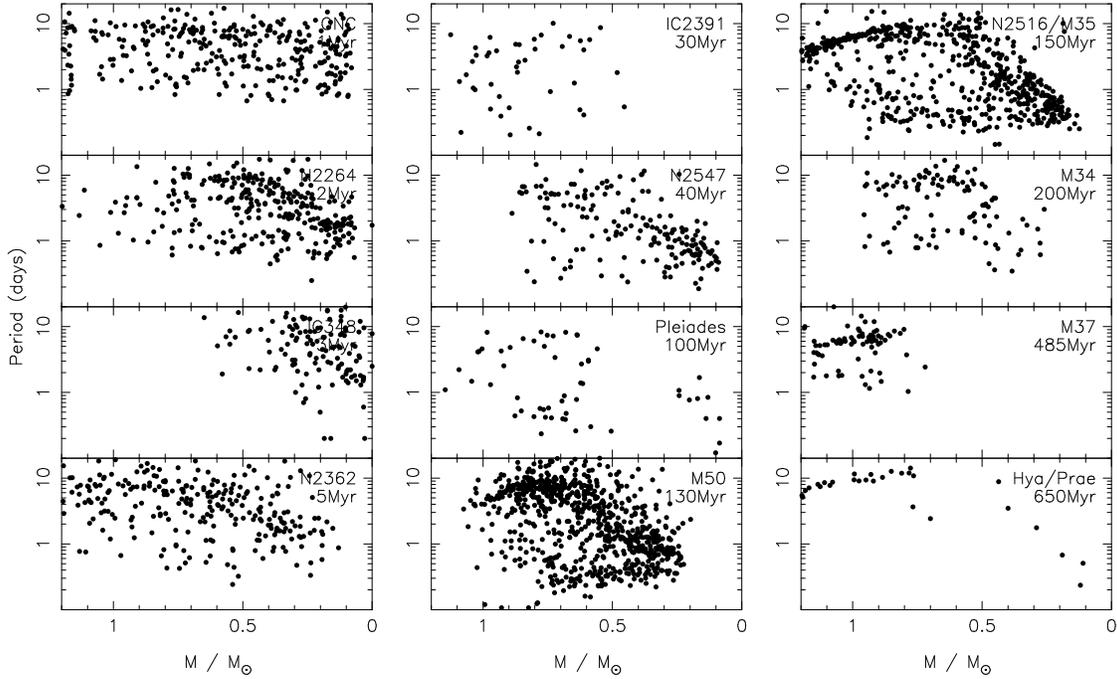}
\caption{Compilation of $\sim 3000$ rotation period and $v\,\sin{i}$
  measurements for open clusters at ages $1-650$\,Myr and masses $<
  1.2$M$_{\odot}$. Masses were derived assuming cluster ages given in
  each panel. Distance and reddening are from the literature using
  $I$-band absolute magnitudes and NextGen stellar models of
  \citet{bcah98}.  For the (numerous) appropriate references, please
  see \citet{i2007}.}
\label{period-mass-comp}
\end{figure}

Rotational evolution has traditionally been constrained by using
measurements in open clusters to provide ``snapshots'' during the
evolution.  Large samples of these are now available covering $\sim
1-650\ {\rm Myr}$ (see Fig.~\ref{period-mass-comp}).  It is becoming
increasingly clear that the evolution is strongly mass and rotation
rate dependent, and this has important consequences for the nature of
the mechanisms governing the angular momentum losses.

In particular, unlike solar-type dwarfs, low-mass dwarfs spin up much
more rapidly, appear to experience no significant angular momentum
losses on the pre main sequence, and much weaker losses due to winds
on the main sequence.

\subsection{Field stars}

Early stars with no relevant convective envelopes cannot generate
surface magnetic fields, they rotate rapidly during their entire
lifetime. Solar-type stars with convective envelopes are strongly
braked as seen above. This is consistent with observations of coronal
emission, chromospheric emission, and latitudinal differential
rotation, which set in exactly where stars are believed to form
convective envelopes, i.e., around spectral type A7--F0. Wind braking
becomes very efficient around late-F type stars, and the Sun for
example has slowed down to less than 2\,km\,s$^{-1}$ during its
lifetime.

Field stars of spectral type K and early-M typically rotate very
slowly as well, although in their youth braking was probably somewhat
weaker (see above). Virtually all single field K and M dwarfs,
including early-M classes M0--M3, are rotating at velocities slower
than about 3\,km\,s$^{-1}$. Around spectral type M3.5, however, a
dramatic increase in rotation rate is observed. This threshold
coincides with the mass range where stars become fully convective. It
appears that for some reason rotational braking becomes weaker at this
boundary.

\begin{figure}
  \includegraphics[height=0.245\textheight]{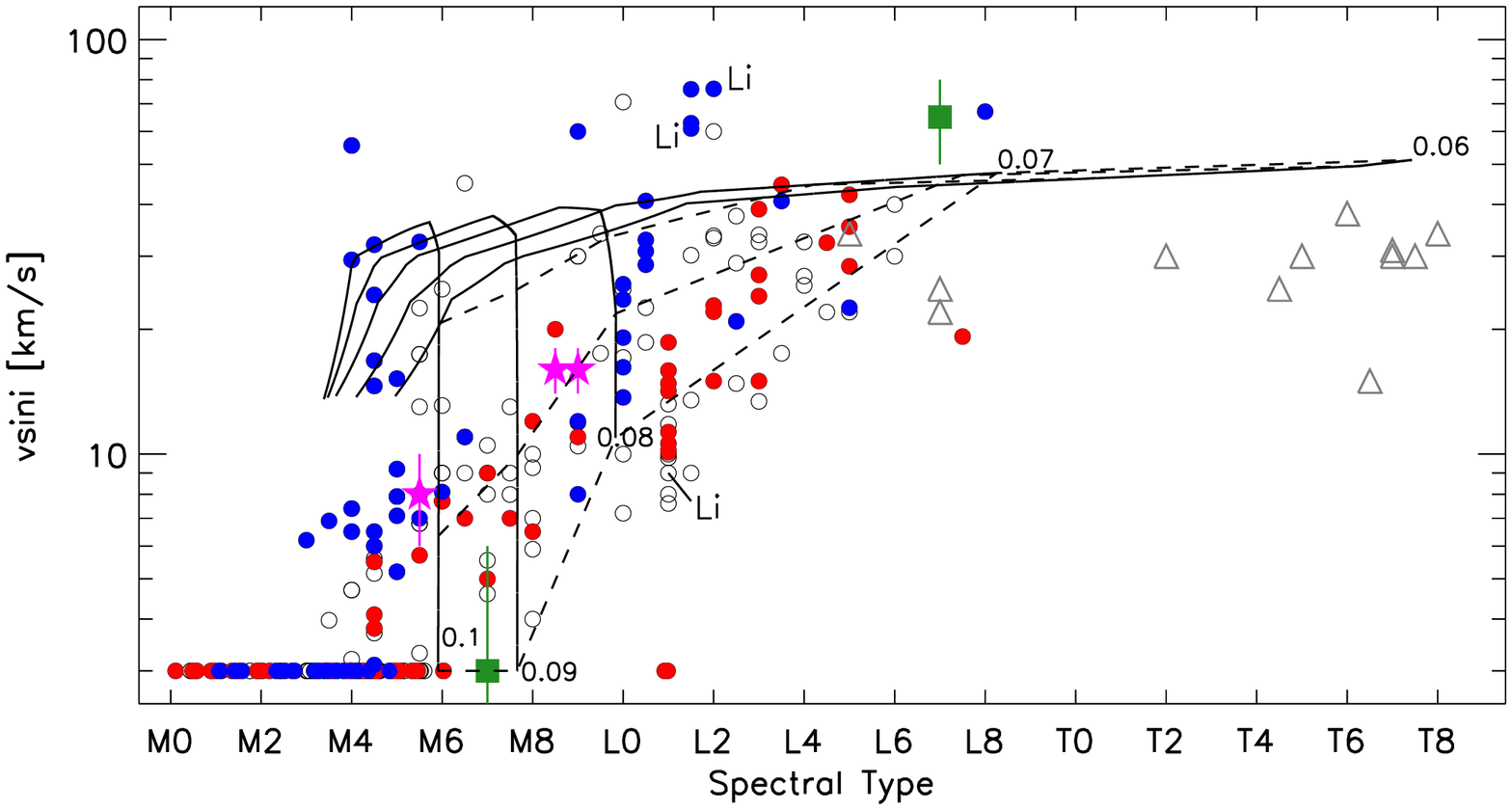}
  \includegraphics[height=0.245\textheight,bbllx=80,bblly=-16,bburx=478,bbury=450]{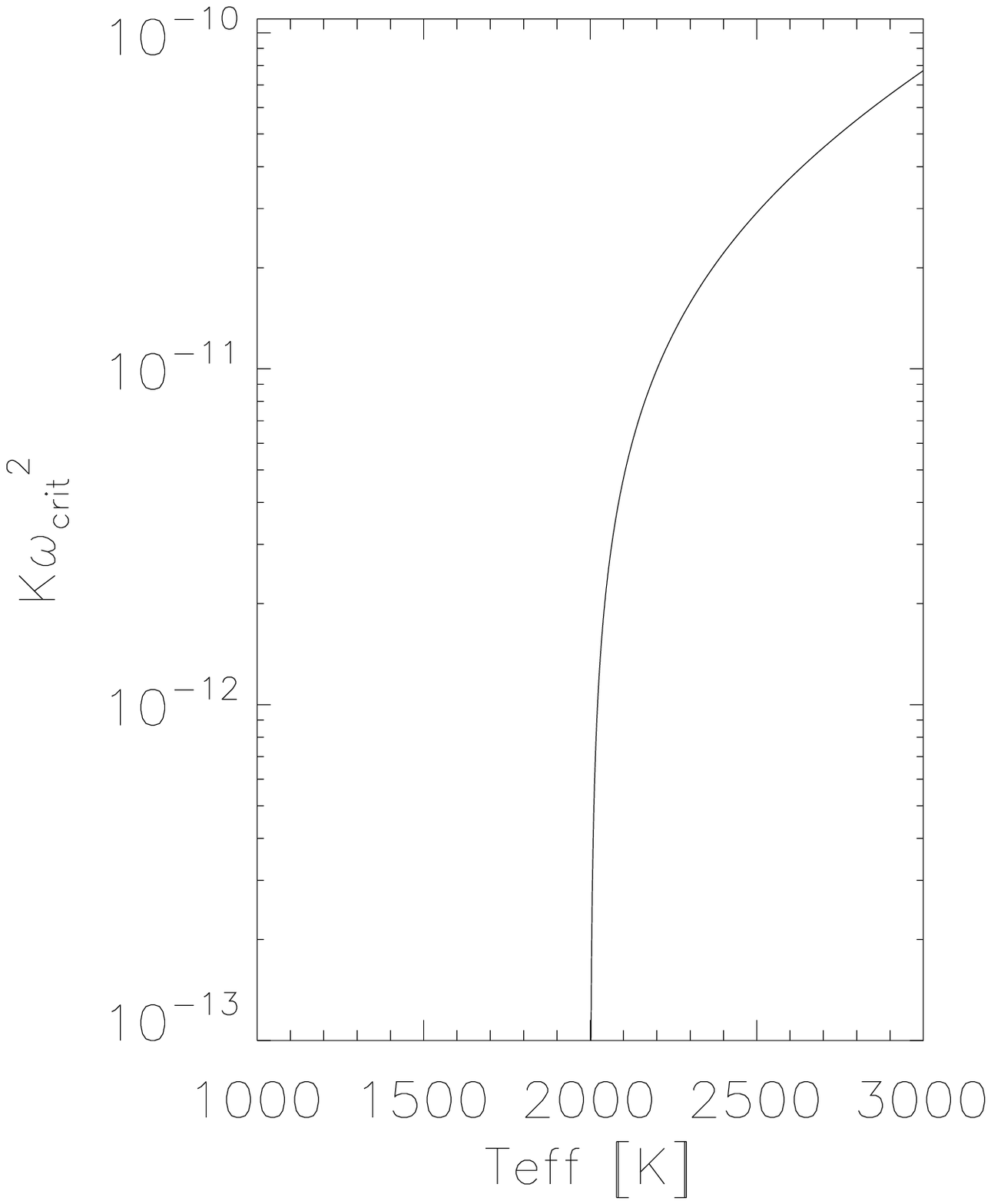}
  \caption{\label{fig:fieldvsini}\emph{Left:} Rotational velocities of
    M--T type objects.  Circles are from \citet{RB07, RB08,
      Delfosse98, Mohanty03} (blue: kinematically young, red:
    kinematically old); triangles from \citet{Zapatero06}.  Magenta
    stars indicate the three members of LHS~1070 \citep{RS07}, filled
    green squares the two subdwarfs 2MASS~0532$+$8246 and
    LSR~1610$-$0040 \citep{RB06a}. Solid lines mark evolutionary
    tracks for objects of 0.1, 0.09, 0.08, and 0.07\,M$_{\odot}$,
    dashed lines mark ages of 1 and 10\,Gyrs (from upper left to lower
    right).  \emph{Right:} Scaling of the magnetic wind-braking with
    temperature in Eq.\ref{eq:windbrake}. }
\end{figure}

Fig.\,\ref{fig:fieldvsini} shows a compilation of (projected)
rotational velocities $v\,\sin{i}$ in objects of spectral classes M--T.
The sudden increase of rotation rate is evident at spectral type
$\sim$M3.5.  Another important result is that braking does not
completely vanish at least until spectral class L0. In the figure,
members of the (statistically) young population are shown in blue and
old stars in red, and the two subdwarfs shown with green squares are
probably \emph{very} old. Young objects are found predominantly in the
upper part of the plot while the old sample shows slower rotation.
This indicates that rotational braking still works in ultra-cool
dwarfs.  Solid lines in Fig.\,\ref{fig:fieldvsini} show evolutionary
tracks according to a modified braking law of the form

\begin{equation}
  \label{eq:windbrake}
  \frac{dJ}{dt} = -K \omega_{\rm crit}^2 \omega \left(\frac{R}{R_\odot}\right)^{0.5} \left(\frac{M}{M_\odot}\right)^{-0.5}.
\end{equation}

Here, $K \omega_{\rm crit}^2$ was scaled according to the right panel
in Fig.\,\ref{fig:fieldvsini}; braking is weaker at lower temperature
\citep[see][]{RB08}.  A viable explanation for this may be the weaker
coupling of magnetic field lines (which still exist) to the atmosphere
that is becoming more and more neutral. Rotational braking in fully
convective field stars and brown dwarfs appears to be so weak that
after a few billion years the distribution of rotational velocities
can tell a lot about their angular momentum evolution and the
underlying processes, magnetic fields and (sub)stellar winds.

\section{Magnetic Fields}

{ The discovery of X-ray emission from the brown dwarf LP944-20
  \citep{Rutledge00} provided the first direct demonstration of
  magnetic activity in the substellar regime.  Subsequent X-ray,
  H$\alpha$, and radio observations revealed that low-mass stars and
  even brown dwarfs ubiquitously generate magnetic activity. No break
  is observed at the boundary to full convection, but chromospheric
  activity weakens after spectral type about M7 \citep{Mohanty03,
    West04, Schmidt07, RB08}, an effect that may be due to decreasing
  fractional ionization \citep{Mohanty02}. Quiescent activity and
  flaring are still observed in even cooler objects
  \citep[e.g.,][]{Hall02, Liebert03, RB08, Robrade08}. About 10\% of
  ultracool dwarfs in the range M7--L4 produce both quiescent and
  flaring radio emission, with inferred field strengths of 0.1--3 kG
  and covering fractions of order unity \citep{Berger06}, and it
  likely correlates with rotation velocity \citep{Berger08b}.  At the
  same time, the tight radio/X-ray correlation that exists in a wide
  range of stars (including the Sun) is strongly violated beyond M7,
  roughly the same regime where chromospheric and coronal emission
  become weaker.  Equally important, several ultracool dwarfs have
  been observed to produce periodic radio emission and H$\alpha$
  emission.  This emission may carry information about the field
  topology.  In general, radio observations suggest that a low
  multipole, large-scale field configuration is the best explanation
  for the observed variability \citep{Berger08a, Hallinan08}.  }

Activity indicators like X-ray, H$\alpha$, and radio emission provide
strong constraints on the magnetic flux depending on the mechanism
that generates the observed emission. Direct measurements of magnetic
fields in M dwarfs through Zeeman splitting of atomic lines were
carried out by \citet{JKV96}, results from a re-analysis with a
multi-component fit are given in \citet{JKV00}. In late-M dwarfs,
however, atomic lines become rare and more and more blended so that
molecular Zeeman diagnostics would be useful, and \citet{Valenti01}
suggested that FeH could be a good indicator of magnetic flux.
\citet{RB06b, RB07} developed a method to measure magnetic flux
through FeH and did so in a sample of M3--M9 dwarfs. They found that
the relation between magnetic fields and (chromospheric) activity is
intact through the entire M spectral range; the most active M stars
exhibit magnetic fields on the order of a few kG. Thus, the lack of
rotational braking in mid- to late-M dwarfs cannot be a consequence of
weaker magnetic fields. Fully convective stars obviously find a way to
efficiently generate magnetic fields.

{
  \section{Magnetic Fields and Wind Braking}
  
  How does the magnetic field connect to rotation?  When a rotating
  star drives an outflow that is well-coupled to the stellar magnetic
  field, the wind and magnetic field conspire to extract angular
  momentum from the star.  This happens because, as wind material
  leaves the stellar surface and tries to conserve its own angular
  momentum, it lags behind the star in a rotational sense.  Thus, the
  magnetic field connecting the stellar surface to the outflowing wind
  is bent backwards with respect to the stellar rotation.  This
  imparts a torque, which acts to give "extra" specific angular
  momentum to the wind, removing it from the star.

  A method for calculating this stellar wind torque dates back to
  \citet{Weber67} and \citet{Mestel68}, and magnetic stellar wind
  theory is still an active research topic.  A generic result is that
  the torque can be written $\tau = \dot M_{\rm w} \Omega_* r_{\rm
    A}^2$, where $\dot M_{\rm w}$ is the mass loss rate in the wind,
  $\Omega_*$ is the angular spin rate of the star, and $r_{\rm A}$ is
  sometimes called the ``magnetic lever arm'' in the flow.  In a
  one-dimensional flow, $r_{\rm A}$ is the Alfv\'en radius, the radial
  location where the wind flow speed equals the magnetic Alfv\'en wave
  speed.

  We can quantify the efficiency of angular momentum extraction by
  dividing the stellar angular momentum by $\tau$, which gives a
  characteristic spin-down time
  \begin{equation}
    \label{eqn_tsd}
    \\
    t_{\rm sd} = k^2 \left({R_* \over r_{\rm A}}\right)^2
    \left({M_* \over \dot M_{\rm w}}\right),
    \\
  \end{equation}
  where $k$ is the ``mean radius of gyration'' (in main sequence
  stars, typically $k^2 \sim 0.1$) and $R_*$ and $M_*$ are the stellar
  radius and mass.  Note that the first two terms on the
  right-hand-side are dimensionless.  The last term has the units of
  time and represents the mass loss time for the star.  In the solar
  wind, for example, $r_{\rm A} / R_* \sim 10$ \citep[e.g.,][]{Li99}.
  Thus the angular momentum loss in magnetic stellar winds can be very
  efficient in a sense that the spin-down time can be much shorter
  than the mass loss time.

  This is an elegant result, but the difficulty lies in calculating
  the effective $r_{\rm A}$ for an arbitrary star and a realistic
  (3-dimensional) wind.  Our understanding of the observed evolution
  of stellar spins depends on this calculation of the torque.  Recent
  work by \citet[][and see contribution in these proceedings]{Matt08}
  emphasizes that, while there is still no adequate theory for
  predicting how the wind torque depends on stellar mass and age,
  significant progress can be made with the use of numerical
  simulations.  }

\section{Stellar Dynamos}

\subsection{Overview}

{ The solar activity cycle is believed to be the result of a dynamo
  process either in the convection zone or the stably stratified layer
  beneath it. The original model was an {$\alpha \Omega$}\ dynamo in
  the convection zone generating a predominantly toroidal and
  axisymmetric magnetic field. Problems with flux storage and the
  internal rotation pattern found by helioseismology led to a revised
  model where the dynamo is located at the bottom of the convection
  zone. That sort of dynamo, however, produces too many toroidal field
  belts and too short cycle periods. The advection-dominated dynamo is
  an extension of the {$\alpha \Omega$}\ dynamo where a large-scale
  meridional flow advects the magnetic field towards the poles at the
  surface and towards the equator at the bottom of the convection
  zone. The butterfly diagram is now the result of the meridional flow
  rather than a dynamo wave and the cycle time depends on the flow as
  much as on the dynamo number.

  \begin{figure}
    \includegraphics[height=.3\textheight]{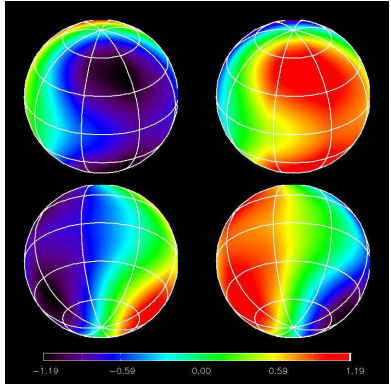}
    \caption{Radial component of a magnetic field of geometry type S1.}
  \end{figure}

  For stars there is no clear picture yet. One would expect stars
  similar to the Sun to show the same type of activity but Doppler
  imaging frequently finds large spots at high latitudes and both
  solar-type and anti-solar cycles have been found in stellar
  butterfly diagrams from photometry. Large polar spots can be
  explained as the consequence of flux tube instability in the
  tachocline while anti-solar butterfly diagrams could indicate a
  meridional flow pattern opposite to that of the Sun.

  Main sequence stars with masses below $\sim0.3$\,M$_\odot$ are fully
  convective, ruling out any dynamo mechanism involving the
  tachocline, but some sort of dynamo must still be at work. The
  $\alpha^2$ dynamo, where the $\alpha$ effect alone generates the
  field, is a possible mechanism. It generates completely
  non-axisymmetric fields that do not oscillate, so that monitoring of
  active low-mass stars will provide an important step towards
  understanding of the dynamo in these stars. At the moment,
  observations support neither the {$\alpha \Omega$}\ nor the
  $\alpha^2$ dynamo: AB~Dor shows pronouced differential rotation but
  a strongly non-axisymmetric surface field while V374~Peg has an
  axisymmetric dipole geometry despite nearly rigid surface rotation
  \citep{Donati06}. }

\subsection{Fully convective stars}

{ Particularly puzzling for dynamo theorists has been the finding that
  fully convective M dwarfs can host \emph{large-scale} magnetic
  fields, even in the absence of any apparent differential rotation.
  \citet{Browning08} discussed 3-D simulations of convection and
  dynamo action in fully convective stars, with an eye toward
  answering two main questions: first, how large-scale fields might be
  generated without a "tachocline" of shear, and second, whether
  differential rotation is always absent in such stars or might be
  maintained in certain circumstances.  In this model
  \citep{Browning08}, convection acted effectively as a dynamo,
  quickly building magnetic fields that (in stars rotating at the
  solar angular velocity) were approximately in equipartition with the
  turbulent velocity field.  More rapidly rotating stars built
  somewhat stronger fields, whereas slower rotators hosted weaker
  fields.  Although differential rotation was established in
  hydrodynamic simulations, the strong magnetic fields realized in
  most MHD cases acted to strongly quench those angular velocity
  contrasts.  Despite the absence of any significant shear, the
  magnetic fields realized in the simulations had structure on a broad
  range of spatial scales, and included a substantial large-scale
  component.  The large-scale field generation is attributed partly to
  the strong influence of rotation upon the slowly overturning flows
  realized in M-stars.
}

\section{Summary}

Our current picture of magnetic field generation, rotation, and
stellar activity may be summarized as follows:

\begin{enumerate}
\item Rotation rates are available for a wide range of masses and
  ages.  Measurements of \emph{projected} rotation velocities extend
  far into the brown dwarf regime, but direct measurements of
  rotational periods are lacking at very low masses.
\item We observe a sharp break in rotation around the threshold where
  stars become fully convective.  This probably indicates a breakdown
  of wind braking.
\item Magnetic field measurements as well as activity tracers like
  X-rays, H$\alpha$, and radio emission show now obvious break at the
  convection boundary.  However, around spectral type M7 normalized
  activity strongly weakens and the relation between radio and X-ray
  emission breaks down.
\item Apparently, very low mass stars can have strong large-scale
  magnetic fields yet only little wind braking. This remains an
  unresolved problem.
\item A key for understanding spindown is a theoretical understanding
  of wind braking.  However, it is still a challenge for magnetic
  stellar wind theory to reliably calculate the wind torque for a
  range of stellar parameters.  Furthermore, the wind torque is
  affected by the mass loss rate, so it is very important that we get
  measurements of mass loss rates and continue to improve mass loss
  theory.
\item Efforts to theoretically understand magnetic field generation
  evolved from the solar dynamo to the larger class of stellar
  dynamos, in particular to fully convective ones in absence of a
  tachocline.  First models successfully reproduce magnetic field
  generation, but it certainly is still a long way to understanding
  magnetic dynamos in very cool stars.
\end{enumerate}


\begin{theacknowledgments}
  We thank the organizers of CS15 for giving us the opportunity to
  hold this session.
\end{theacknowledgments}

\end{document}